\documentclass[aps,prl,twocolumn,showpacs]{revtex4}
\usepackage{graphicx}

\begin{document}
\title{Near-Perfect Correlation of the Resistance Components of Mesoscopic Samples at the Quantum Hall Regime 
}
\author{E. Peled$^{1}$}
\author{D. Shahar$^{1}$}
\author{Y. Chen$^{2}$}
\author{E. Diez$^{2}$}
\thanks{Present address: Departamento de F\'{i}sica, Facultad de Ciencias, Universidad de Salamanca, E-37008 Salamanca, Spain}
\author{D. L. Sivco$^{3}$}
\author{A. Y. Cho$^{3}$}
\affiliation{
$^{1}$Department of Condensed Matter Physics, The Weizmann Institute of Science, Rehovot 
76100, Israel,\\
$^{2}$Department of Electrical Engineering, Princeton University, 
Princeton, New Jersey 08544, USA,\\
$^{3}$Bell Laboratories, Lucent Technologies, 600 Mountain Avenue, Murray Hill, 
New Jersey 07974, USA.
}
\date{\today}

\begin{abstract}
We study the four-terminal resistance fluctuations of mesoscopic samples near the transition between the $\nu=2$ and the $\nu=1$ quantum Hall states. We observe near-perfect correlations between the fluctuations of the longitudinal and Hall components of the resistance. These correlated fluctuations appear in a magnetic-field range for which the two-terminal resistance of the samples is quantized. We discuss these findings in light of edge-state transport models of the quantum Hall effect. We also show that our results lead to an ambiguity in the determination of the width of quantum Hall transitions. 
\end{abstract}


\maketitle

     
When placed in strong magnetic fields ($B$s), two-dimensional electron systems can display a series of states known as the Quantum Hall effect (QHE) \cite{Klitzing1980PRL45}. These states display a remarkable universality:
irrespective of many of the system's properties such as geometry and disorder strength, its  Hall resistance ($R_{H}$) exhibits exact quantization at $h/ie^{2}$ ($h$ is Planck's constant, $e$ is the charge of the electron, and $i$ is an integer), while its longitudinal resistance ($R_{L}$) vanishes. 

When the size of the samples becomes smaller, approaching the mesoscopic regime, the features of the QHE begin to diminish. In addition, a pattern of reproducible fluctuations appears, whose magnitude increases as the sample size and the temperature ($T$) decrease. Near $B=0$ these are the well-known universal conductance fluctuations \cite{Lee1987PRB35} famous for the universality of their amplitude, which is close to $e^2/h$. In the quantum Hall (QH) regime the understanding of the fluctuations is not as complete, despite the large number of 
experimental \cite{Timp1987PRL59, Chang1988SSC67, Staring1992PRB45, Main1994PRB50, Machida2000PE6, Cobden1996PRB54, Cobden1999PRL82, Machida2001PRB63, Simmons1989PRL63, Goldman1995SC267, Geim1992PRL69, Morgan1994PRB50, Hohls2002PRB66} and theoretical 
\cite{Ando1994PRB49, Wang1996PRL77, Galstyan1997PRB56, Cho1997PRB55, Ho1999PRB60, Jain1988PRL60, Maslov1993PRL71, Xiong1997PRB56} studies. In particular, the amplitude of the fluctuations in this regime shows a distinct $B$ dependence and is not universal.

In this Letter we report on the observation of a new type of universal behavior of the fluctuations in the QH regime. We have measured $R_{L}$ and $R_{H}$ under the conditions of the QHE, in mesoscopic samples for which finite-size effects are dominant. Our samples display reproducible resistance-fluctuations that are cool-down, as well as contact configuration, specific. We found that there are near-perfect correlations between the fluctuations measured in $R_{L}$ and those measured in $R_{H}$. Specifically, in the vicinity of the transition between the $\nu=2$ and the $\nu=1$ QH states, we find that 
\begin{equation}
    R_{L}+R_{H}=h/e^{2}
\label{eq:Correlation}
\end{equation}
over a wide range of $B$. We trace the origin of these correlations to the quantization, over the same $B$-range, of the two-terminal resistance of the sample ($R_{2t}$). The link between the sum $R_{L}+R_{H}$ and $R_{2t}$ is in accordance with the transport model of Streda {\it et al.}\ \cite{Streda1987PRL59}, that combines the Landauer formulation for conductance with the existence of electronic edge states at high $B$s \cite{Buttiker1988PRB38, McEuen1990PRL64}. Finally, we demonstrate that our findings reveal an ambiguity in the determination of the width of QH transitions, a property that is material to the subject of scaling and universality in QH transitions.

\begin{figure}[ht]
\includegraphics[scale=0.55]{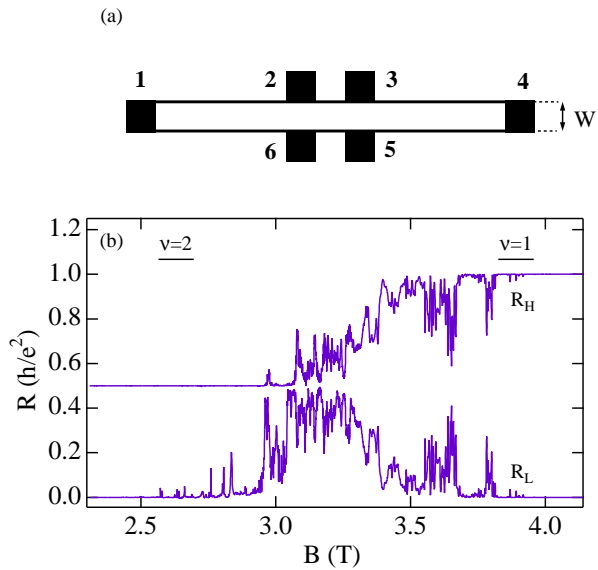} 
\caption{
(a) Geometry of the Hall-bar samples. The black areas represent Au-Ge-Ni contacts. The separation of the current and voltage contacts are $12 \times W$  and $2 \times W$, respectively. 
(b) $R_{L}$ and $R_{H}$ vs $B$ of the $2\ \mu$m Hall-bar in the vicinity of the $\nu=$2-1 transition, $T=10$ mK.
}
\label{fig:Fig1}
\end{figure}

The samples we used (T2Cm2, T2Cm20, and T2Cm100) were made from a InGaAs/InAlAs wafer containing a 200 \AA\ quantum well. The short-range scattering in the wafer leads to the formation of a low-mobility, low-density two-dimensional electron system, after illumination with an LED. Our samples have an average density $n_{s}=1.15 \cdot 10^{11}$ cm$^{-2}$ and average mobility $\mu = 16,600$ cm$^{2}$/Vsec, limiting our study to the integer QHE. 
We have defined three Hall-bar samples, wet-etched with the same aspect ratio (see Fig.\ \ref{fig:Fig1}(a)), but with lithographic widths of $W=$ 2, 20, and 100 $\mu$m. To ensure maximum uniformity, the three samples were prepared on the same chip within 2 mm of each other. The black areas in Fig.\ \ref{fig:Fig1}(a) represent Au-Ge-Ni alloyed contacts that were designed to reach the edges of the Hall-bars. The samples were cooled in a dilution refrigerator with a base $T$ of 10 mK, at which all of the data presented here were taken. Four- and two-terminal measurements were done using standard AC lock-in techniques with a frequency of 3.17 Hz and an excitation current of 1 nA. The value of 1 nA for the current was chosen to avoid electron heating. At higher current values ($I \geq$ 10 nA) we find evidence for heating: the resistance fluctuations diminish in magnitude and the width of the QH transitions increases. 

We begin the description of our data by presenting, in Fig.\ \ref{fig:Fig1}(b), $B$ traces of $R_{L}$ and $R_{H}$ for the 2 $\mu$m Hall-bar in the vicinity of the transition between the $\nu=2$ and the $\nu=1$ QH states. Referring to the diagram in Fig.\ \ref{fig:Fig1}(a), and using the standard notation $R_{ij,kl}=V_{kl}/I_{ij}$, where $V_{kl}$ is the voltage difference between probes $k$ and $l$ and $I_{ij}$ is the current between probes $i$ and $j$, the data we show are $R_{L}=R_{14,65}$ and $R_{H}=R_{14,53}$.
Despite the small size of the sample, the $\nu=1$ ($B>3.95$ T) and $\nu=2$ ($B<2.55$ T) QH states are clearly seen, evident by the quantization of $R_{H}$ and the corresponding vanishing of $R_{L}$. 

The finite size of the sample is manifested by the appearance, in the transition region, of large, reproducible, fluctuations in both $R_{L}$ and $R_{H}$. As seen in previous studies of mesoscopic samples in the QH regime,
these noise-like fluctuations maintain their pattern as long as the sample is kept cold, and diminish in magnitude as $T$ is increased. A new fluctuation pattern is found each time the sample is temperature-cycled. 

The central finding of our work is the existence, on the $\nu=1$ side of the transition ($B=$ 3.1 -- 3.9 T in Fig.\ \ref{fig:Fig1}(b)), of near-perfect correlations between the fluctuations of $R_{L}$ and those of $R_{H}$.
Graphically, we observe that for each peak in $R_{L}$ there corresponds a dip in $R_{H}$ of nearly equal magnitude, and vice versa. 
This holds for almost all the fine details of the fluctuation patterns. While such correlations could arise from mixing of the resistance components, it is unlikely that this is the case in our work since the correlations are limited to a specific range of $B$ and do not show up at either low $B$ or at the $\nu=2$ side of the transition. 

\begin{figure}[ht]
\includegraphics[scale=0.55]{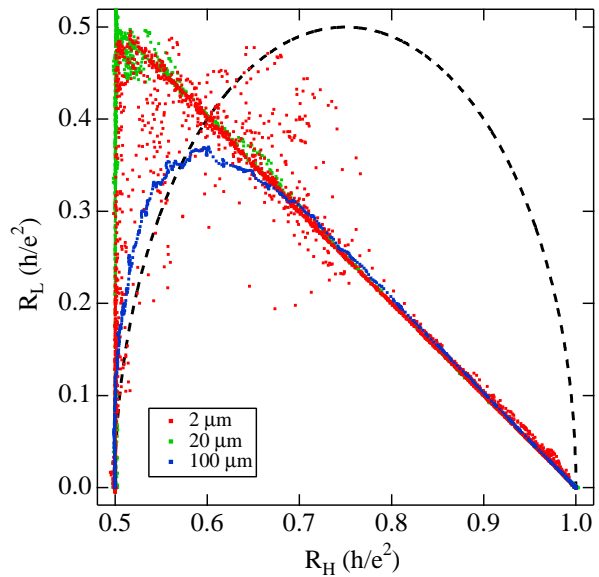} 
\caption{
$R_{L}$ vs $R_{H}$ for the 2, 20, and 100 $\mu$m Hall-bars in the vicinity of the $\nu=$2-1 transition, $T=10$ mK. Dashed line: the theoretical semicircle relation for a macroscopic sample with the same aspect ratio as our samples.
}
\label{fig:Fig2}
\end{figure}

Mathematically, the correlations we observe can be conveniently expressed by the simple relation of Eq.\ \ref{eq:Correlation}. 
To see this dependence more clearly we plot, in Fig.\ \ref{fig:Fig2}, $R_{L}$ vs $R_{H}$ for all three samples. In this unconventional plot \cite{McEuen1990PRL64} each dot represents one ($R_{H}$, $R_{L}$) data-pair from $B$ traces such as those in Fig.\ \ref{fig:Fig1}(b). Focusing on the data from the 2 $\mu$m sample of Fig.\ \ref{fig:Fig1}(b) (red dots in Fig.\ \ref{fig:Fig2}) we see that,
aside from some scatter, the dots fall into two ordered groups: a diagonal line stretching from ($0.5\ h/e^{2}$, $0.5\ h/e^{2}$) to ($h/e^{2}$, $0$) and a vertical line at $R_{H}=0.5\ h/e^{2}$. The diagonal line corresponds to $R_{L}+R_{H}=h/e^{2}$, and is comprised of the correlated ($R_{H}$, $R_{L}$) data-pairs from the $\nu=1$ side of the transition. The dots that form the vertical line are from the $\nu=2$ side of the transition, in the $B$ range of 2.6 -- 2.9 T in Fig.\ \ref{fig:Fig1}(b). In that $B$ range $R_{H}$ remains quantized at 0.5 $h/e^{2}$, while $R_{L}$ can take any value in the range 0 -- 0.5 $h/e^{2}$. 
The remaining, scattered, dots are mainly from the intermediate $B$ range (2.9 -- 3.1 T in Fig. \ref{fig:Fig1}(b)) of the transition between the QH states, and also include the (relatively few) deviations from the ordered lines. We note that the observed $R_{L}$-$R_{H}$ relation is different from the derivative-law relating the resistivity components observed in macroscopic samples \cite{Chang1985SSC56, Rotger1989PRL62, Stormer1992SSC84, Simon1994PRL73}.
 
Fig.\ \ref{fig:Fig2} also includes data obtained from the 20 and 100 $\mu$m Hall-bars (green and blue dots, respectively). While the 20 $\mu$m dots exhibit similar behavior to those of the 2 $\mu$m sample, the 100 $\mu$m sample shows somewhat different characteristics. When $R_{H}>0.7\ h/e^{2}$ the 100 $\mu$m dots are close to the $R_{L}+R_{H}=h/e^{2}$ diagonal line, but otherwise they form a continuous curve, with $R_{L}$ values that are lower than the corresponding 2 and 20 $\mu$m $R_{L}$ values, and do not split into either a diagonal or vertical line. We attribute this difference in the $R_{L}$-$R_{H}$ dependence to the larger size of the 100 $\mu$m Hall-bar. An infinite, homogenous, sample with the same aspect ratio as our samples is expected to have a semicircle $R_{L}$-$R_{H}$ dependence as shown in dashed line in the figure \cite{Ruzin1995PRL74, Semicircle}. Comparing the measured data with the semicircle trace we find that although the 100 $\mu$m Hall-bar has the characteristics of a wider sample, it may not be large enough to exhibit the full semicircle behavior. This may be related to the fact that resistance fluctuations in the 100 $\mu$m Hall-bar begin to be discernible at our lowest $T$. 

The clear ordering of the ($R_{H}$, $R_{L}$) pairs evident in Fig.\ \ref{fig:Fig2}, and the fact that data from different size samples fall on top of each other, are surprising from several respects. First, the $\nu=$ 2-1 transition does not take place at the same $B$ range in all samples, due to small differences in electron density. Second, the apparent $B$-width of the transitions, although not clearly defined due to the large fluctuations present in the 2 and 20 $\mu$m samples, varies between samples of different widths and is larger for the narrower samples (see Fig.\ \ref{fig:Fig4} and discussion below) \cite{Koch1991PRL67, Machida1996PRB54}.
And third, the random nature of the fluctuations, unique to each sample and cool-down, indicates that a fundamental mechanism underlies the appearance of order in the data of Fig.\ \ref{fig:Fig2}.

The theoretical model that is most suitable for discussing transport in mesoscopic samples in the QH regime is the edge-state model \cite{McEuen1990PRL64, Chang1989SSC72, Haug}. In this model the electrons move along one-dimensional channels that follow the edges of the sample, with the direction of their motion set by the polarity of $B$. The resistance of the sample can be determined, following the Landauer formulation, by the probabilities of an edge-state electron to be transmitted forward along the same edge or reflected to a different edge of the sample. Using this approach, Streda {\it et al.}\ \cite{Streda1987PRL59} and B\"{u}ttiker \cite{Buttiker1988PRB38} were able to derive explicit formulas for the resistances in the QH regime.

\begin{figure}[ht]
\includegraphics[scale=0.55]{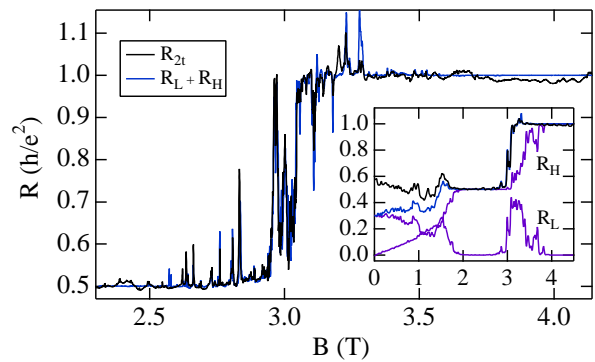} 
\caption{
$R_{2t}$ and $R_{L}+R_{H}$ of the $2\ \mu$m Hall-bar in the vicinity of the $\nu=$2-1 transition, $T=10$ mK. The $R_{2t}$ trace is shown after a subtraction of 1,402 $\Omega$. Inset: $R_{L}$ and $R_{H}$ (purple), $R_{2t}$ and $R_{L}+R_{H}$ over a wider $B$ range. Note that $R_{L}+R_{H} \neq R_{2t}$ for $B<2$ T.
}
\label{fig:Fig3}
\end{figure}

\begin{figure}[ht]
\includegraphics[scale=0.55]{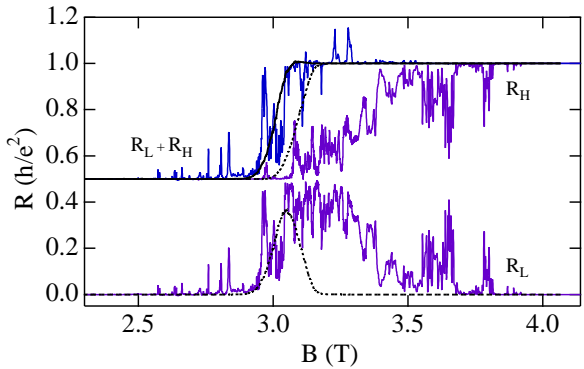} 
\caption{
$R_{L}$ and $R_{H}$ (purple) and $R_{L}+R_{H}$ (blue) of the 2 $\mu$m Hall-bar, together with $R_{L}$ and $R_{H}$ (dashed black) and $R_{L}+R_{H}$ (solid black) of the $100\ \mu$m  Hall-bar, in the vicinity of the $\nu=$2-1 transition, $T=10$ mK.
The 100 $\mu$m traces are shifted to the left by 0.074 T.
}
\label{fig:Fig4}
\end{figure}

An intriguing result that directly stems from the Landauer analysis of QH samples was pointed out by Streda {\it et al.} \cite{Streda1987PRL59}. They explicitly calculated $R_{L}$ and $R_{H}$ and showed that they obey the simple sum rule  \cite{Buttiker}
\begin{equation}
    R_{L}+R_{H}=R_{2t}.
\label{eq:SumRule}
\end{equation}
To test this prediction we plot, in Fig.\ \ref{fig:Fig3}, $R_{2t}$ ($R_{63,63}$, black line) of the 2 $\mu$m Hall-bar together with the sum $R_{L}+R_{H}$ (blue line) of the resistances from Fig.\ \ref{fig:Fig1}(b). $R_{2t}$ is plotted after subtracting a $B$-independent contact resistance of 1,402 $\Omega$, chosen by requiring that $R_{2t}$ will be equal  to $R_{H}$ deep in the $\nu=$ 1 QH state. 
As can be seen, the agreement between our data and Eq.\ \ref{eq:SumRule} is very good, and includes the overall shape of the resistance trace between the $\nu=2$ to the $\nu=1$ QH states as well as most of the fluctuations.

The simple sum rule expressed by Eq.\ \ref{eq:SumRule}, together with its verification in Fig.\ \ref{fig:Fig3}, may seem, at first glance, a natural consequence of Kirchhoff's law. This becomes clear if we rewrite  Eq.\ \ref{eq:SumRule} as
$(V_{65}+V_{53})/I_{14}=V_{63}/I_{14}=V_{63}/I_{63}$, 
and remember that we use the same value of current, $I_{14}=I_{63}=1$ nA, for both measurements. However, we must keep in mind that the current paths, and the measurement geometry, in the two measurement configurations are different, and therefore the second equality in the equation above should not hold. Wider $B$-range measurements of our 2 $\mu$m sample, shown in the inset to Fig.\ \ref{fig:Fig3}, indicate that the sum rule of Eq.\ \ref{eq:SumRule} is clearly violated near $B=0$ and also where QH features are not fully developed. We can therefore regard Eq.\ \ref{eq:SumRule} to be a special property of the QH regime.

In Fig.\ \ref{fig:Fig3} we have shown that the simple sum-rule of Eq.\ \ref{eq:SumRule} predicted by Streda {\it et al.}\ \cite{Streda1987PRL59} holds over the entire range of $B$ covering the $\nu=2$ and $\nu=1$ QH states and the transition region between them. We have also shown that, over a limited, but large, $B$-range, our data obey the experimentally derived Eq.\ \ref{eq:Correlation}. The coincidence of the two relations is where $R_{2t}=h/e^{2}$. An intriguing question is why the quantization of $R_{2t}$ is maintained over a much broader range of $B$ than the quantization of the four-terminal $R_H$ (and the vanishing of $R_L$). 

While we are unable to answer this question, we wish to remark on an additional difficulty that arises from this observation. This difficulty is related to the determination of the transition width, along with its $T$ dependence, which are key parameters in the description of QH transitions \cite{Pruisken1988PRL61, Wei1988PRL61, Koch1991PRL67, Machida1996PRB54}. In order to define the transition region for mesoscopic samples one fits a smooth function to the fluctuating data and obtains the width from this fit. If we apply this procedure to $R_H$ and $R_L$ that are obtained from the 2 $\mu$m sample, we find a much broader transition than the corresponding transition in the 100 $\mu$m sample. If, on the other hand, we use $R_{2t}$ or, equivalently,  the combination $R_{L}+R_{H}$ the resulting width is similar to that obtained from the 100 $\mu$m sample. This is illustrated in Fig.\ \ref{fig:Fig4}, where we plot $R_{L}$, $R_{H}$, and $R_{L}+R_{H}$ of the 2 and 100 $\mu$m Hall-bars at the vicinity of the $\nu=$2-1 transition. $R_{L}$ and $R_{H}$ of the 100 $\mu$m Hall-bar display very small fluctuations and their transition region is narrower than that reflected from $R_{L}$ and $R_{H}$ of the 2 $\mu$m Hall-bar. In contrast, for both samples, the sum $R_{L}+R_{H}$ has approximately the same width. 

Generally, in mesoscopic samples, measurements that use different contact configurations yield different average resistance and fluctuation patterns. 
In our samples we find that the different resistance measurements are related in a way that is consistent with the $R_{L}$-$R_{H}$ correlations discussed above. All of the measurements that were presented thus far were done using the contact configuration of $R_{L}=R_{14,65}$ and $R_{H}=R_{14,53}$. The relations to other contact configurations can be summarized as follows: each possible contact configuration of $R_{H}$ ($R_{14,62}$ or $R_{14,53}$) results in a different resistance and fluctuation pattern. To each of these two options there corresponds one correlated $R_{L}$ configuration ($R_{14,23}$ or $R_{14,65}$, respectively): $R_{14,65}+R_{14,53}=R_{14,23}+R_{14,62}=R_{63,63}$. When reversing the $B$ polarity $R_{H}$ changes sign and the corresponding $R_{L}$ configuration is switched to the other side of the Hall-bar: $R_{H}(-B)=-R_{H}(B)$ and $R_{14,23}(\mp B)=R_{14,65}(\pm B)$ \cite{Ponomarenko2003CM0306063}.

To conclude, we have shown that in the QH regime $R_{L}+R_{H}=R_{2t}$, in agreement with the results of the model of edge-state conduction. 
For the $B$ range where $R_{2t}=h/e^{2}$ this leads to $R_{L}$-$R_{H}$ correlations of the form $R_{L}+R_{H}=h/e^{2}$. We have pointed out difficulties in estimating the width of QH transitions in mesoscopic samples.

We wish to thank discussions with Y. Imry, Y. Oreg, A. Stern, and D. C. Tsui. This work is supported by the BSF and by the Koshland Fund. Y. C. is supported by the (U.S.) NSF. E.D. is supported by the Ram\'{o}n y Cajal Program of the Spanish Minister of Science and Technology.


\begin{thebibliography}{10}
\bibitem{Ruzin1995PRL74}
I. Ruzin and S. Feng, Phys. Rev. Lett. {\bf 74},  154  (1995), these authors
  predict a semicircle relation for the conductivities that, upon inversion of
  the conductivity tensor, is also a semicircle for the resistivities.
\end{thebibliography}

\end{document}